\documentclass[10pt, conference]{IEEEtran}
\IEEEoverridecommandlockouts
\usepackage{cite}
\usepackage{amsmath,amssymb,amsfonts}
\usepackage{bbm}
\usepackage{algorithmic}
\usepackage{graphicx}
\usepackage{textcomp}
\usepackage{xcolor}
\usepackage{soul}
\usepackage{siunitx}
\usepackage{subcaption}
\usepackage[printonlyused]{acronym}

\acrodef{mmW}{millimeter-wave}
\acrodef{sub-THz}{sub-terahertz}
\acrodef{CS}{compressive sensing}
\acrodef{CPR}{compressive phase retrieval}
\acrodef{BS}{base station}
\acrodef{UE}{user equipment}
\acrodef{Tx}{transmitter}
\acrodef{Rx}{receiver}
\acrodef{SNR}{signal-to-noise ratio}
\acrodef{AWV}{antenna weight vector}
\acrodef{LOS}{line-of-sight}
\acrodef{NLOS}{non-line-of-sight}
\acrodef{BA}{beam alignment}
\acrodef{BT}{beam tracking}
\acrodef{BP}{beam prediction}
\acrodef{IA}{initial access}
\acrodef{AoA}{angles-of-arrival}
\acrodef{AoD}{angle-of-departure}
\acrodef{RSS}{received signal strength}
\acrodef{MP}{matching pursuit}
\acrodef{ML}{machine learning}
\acrodef{PN}{pseudorandom noise}
\acrodef{QPD}{quadratic phase distribution}
\acrodef{SA}{sub-array}
\acrodef{CNN}{convolutional neural network}
\acrodef{FCnet}{fully-connected neural network}
\acrodef{MLP}{multi-layer perceptron}
\acrodef{NLP}{natural language processing}
\acrodef{DSP}{digital signal processing}
\acrodef{ULA}{uniform linear array}
\acrodef{Seq2Seq}{sequence-to-sequence}
\acrodef{LSTM}{long short-term memory}
\acrodef{RNN}{recurrent neural network}

\usepackage{tikz}
\usepackage{textcomp}
\usepackage{hyperref}
\newcommand\copyrighttext{%
  \footnotesize \textcopyright 2022 IEEE. Personal use of this material is permitted.
  Permission from IEEE must be obtained for all other uses, in any current or future
  media, including reprinting/republishing this material for advertising or promotional
  purposes, creating new collective works, for resale or redistribution to servers or
  lists, or reuse of any copyrighted component of this work in other works. 
}
\newcommand\copyrightnotice{%
\begin{tikzpicture}[remember picture,overlay]
\node[anchor=south,yshift=10pt] at (current page.south) {\fbox{\parbox{\dimexpr\textwidth-\fboxsep-\fboxrule\relax}{\copyrighttext}}};
\end{tikzpicture}%
}

\DeclareMathOperator*{\argmax}{arg\,max}

\def\BibTeX{{\rm B\kern-.05em{\sc i\kern-.025em b}\kern-.08em
    T\kern-.1667em\lower.7ex\hbox{E}\kern-.125emX}}

\begin{document}
\bstctlcite{IEEEexample:BSTcontrol}


\title{Machine Learning Prediction for Phase-less Millimeter-Wave Beam Tracking\\
\thanks{This work is supported by NSF under grant 1718742. This work was also supported in part by the ComSenTer and CONIX Research Centers, two of six centers in JUMP, a Semiconductor Research Corporation (SRC) program sponsored by DARPA.}
}

\author{\IEEEauthorblockN{Benjamin W. Domae, Veljko Boljanovic, Ruifu Li, and Danijela Cabric}
\IEEEauthorblockA{\textit{Electrical and Computer Engineering Department,} \\
\textit{University of California, Los Angeles}\\
Emails: bdomae@ucla.edu, vboljanovic@ucla.edu, doanr37@ucla.edu, danijela@ee.ucla.edu }
}

\maketitle
\copyrightnotice

\begin{abstract}
Future wireless networks may operate at \ac{mmW} and \ac{sub-THz} frequencies to enable high data rate requirements. While large antenna arrays are critical for reliable communications at \ac{mmW} and \ac{sub-THz} bands, these antenna arrays would also mandate efficient and scalable initial beam alignment and link maintenance algorithms for mobile devices. Low-power phased-array architectures and phase-less power measurements due to high frequency oscillator phase noise pose additional challenges for practical beam tracking algorithms. Traditional beam tracking protocols require exhaustive sweeps of all possible beam directions and scale poorly with high mobility and large arrays. Compressive sensing and machine learning designs have been proposed to improve measurement scaling with array size but commonly degrade under hardware impairments or require raw samples respectively. In this work, we introduce a novel \ac{LSTM} network assisted beam tracking and prediction algorithm utilizing only phase-less measurements from fixed compressive codebooks. We demonstrate comparable beam alignment accuracy to state-of-the-art phase-less beam alignment algorithms, while reducing the average number of required measurements over time.
\end{abstract}


\section{Introduction}
Due to large available bandwidth, \ac{mmW} frequency bands are the key candidates for high data rate cellular and wireless local area networks \cite{Andrews:5G}. Their use, however, comes at the cost of a higher propagation loss than at sub-6 GHz frequencies \cite{Rappaport:propagation}. The \ac{BS} and \ac{UE} with large antenna arrays need to establish a directional communication link through a \ac{BA} procedure to compensate for this loss. Additionally, in highly dynamic \ac{mmW} environments, where the channel parameters change quickly, the optimal steering beams at the \ac{BS} and \ac{UE} need to be tracked between two beam training procedures. 
In recent years, the \ac{BT} problem in \ac{mmW} networks have attracted significant attention from researchers. Previous work on \ac{BT} can be roughly divided into two groups, as discussed below.

The first group of work aims to leverage advanced \ac{DSP} techniques and/or array architectures for \ac{BT}, including different algorithms based on compressive sensing \cite{Marzi:compressive,Boljanovic:tracking} and Kalman filtering \cite{Va:tracking, Jayaprakasam:robust_tracking}. Compressive algorithms exploit the sparsity of \ac{mmW} channels, reducing the required tracking overhead by taking a small number of signal measurements. On the other hand, Kalman filtering can be leveraged for continuous \ac{BT}. However, imperfect synchronization and noise can significantly reduce the tracking performance for both the compressive and Kalman filtering methods. \ac{DSP} algorithms based on phase-less power-based measurements can mitigate this sensitivity problem, as \cite{Rasekh2017_CSRSSMP} has done for compressive sensing and \cite{Jain:mMobile} for Kalman filtering.

In the second group of work, a number of \ac{ML} algorithms have been proposed for \ac{mmW} \ac{BT} and \ac{BP}. Previous work has considered various \ac{RNN} architectures to learn and exploit the temporal evolution of the optimal beam steering directions. In \cite{Jiang:rnn_tracking}, the authors proposed an encoder-decoder model with a vanilla \ac{RNN} network to predict multiple future steering directions based on the previously estimated beams and visual sensing information. \cite{Burghal2019_MLTracking} proposed a simple \ac{LSTM} network to estimate the current dominant \ac{AoA}, while \cite{Shah2021_LSTMautoencoderPrediction} presented an \ac{LSTM} with autoencoder feature extraction to predict path \ac{SNR} between multiple \ac{mmW} \ac{BS}s.
Existing work on \ac{ML} models for \ac{BT} and \ac{BP} primarily focus on sample-based techniques, instead of compressive phase-less measurements.

In this work, we propose a novel \ac{RNN} algorithm to reduce \ac{BT} communications overhead for \ac{UE}s with phased-array antennas.
The \ac{ML} architecture is based on a \ac{Seq2Seq} translation model and \ac{LSTM} networks, predicting future directional pencil beams for data communications using a buffer of prior compressive \ac{BA} measurements. To our best knowledge, this is the first work to propose a data-driven \ac{BT} and \ac{BP} algorithm using only phase-less \ac{RSS} measurements from fixed sensing codebooks, addressing practical hardware constraints. Prior \ac{BT} and \ac{BP} algorithms require raw samples or adaptive codebooks.

The remainder of this work is organized as follows: Section \ref{sec:system} discusses our system model, problem statement, metrics, and existing baseline solutions for phase-less, fixed-codebook \ac{BA}. Section \ref{sec:algo_design} details our proposed algorithm design while section \ref{sec:eval_design} presents our simulation design and algorithm performance results. Finally, we conclude in Section \ref{sec:conclusion}.

\textit{Notation}: Scalars, vectors, and matrices are represented by non-bold lowercase, bold lowercase, and bold uppercase letters respectively. The transpose and Hermitian transpose of $\mathbf{A}$ are $\mathbf{A}^T$ and $\mathbf{A}^H$ respectively.

\section{System Model and Problem Statement}
\label{sec:system}

\subsection{Time-Varying Channel Model}
\label{subsec:channel}
We consider a narrowband \ac{mmW} \ac{LOS} channel between the \ac{BS} and \ac{UE}. Due to the \ac{UE}'s mobility, the angle and gain of the \ac{LOS} path may evolve non-linearly over time. We assume that the \ac{BS} is equipped with a digital array, which enables it to quickly estimate and adapt its beam/precoder to the mobile \ac{UE}. Thus, with a $N_r$-element \ac{ULA} at the \ac{UE}, the effective channel vector $\mathbf{h}\in\mathbb{C}^{N_r}$ at the time step $t$ can be expressed as:
\begin{equation}
    \mathbf{h}^{(t)} = e^{-\alpha^{(t)}}\mathbf{a}_{r}\left( \phi_l^{(t)} \right)
    \label{eq:channel}
\end{equation}
where $e^{-\alpha^{(t)}}$ describes the path loss and \ac{BS} antenna gain. The $n$-th element of the \ac{UE}'s spatial response $\mathbf{a}_r(\phi_l^{(t)})$ is defined as $[\mathbf{a}_r(\phi_l^{(t)})]_n = \exp{(j2\pi(n-1)\sin{(\phi_l^{(t)})}d/\lambda)}$, with $\lambda$ and $d$ being the wavelength and separation between the \ac{UE} antennas.

\subsection{Received Signal Model}
\label{subsec:received_signal}
As in \cite{Yan2020_mmrapid}, where a downlink \ac{BA} problem was studied, we assume the \ac{UE} conducts phase-less power measurements using a predefined sensing codebook of \ac{PN} beams. Phase-less measurements have two main advantages over the measurements with phase information: 1) they do not require a precise and complex synchronization between the \ac{BS} and \ac{UE}; 2) they can be more robust to random hardware impairments that are often present in practical arrays, especially when paired with \ac{ML} algorithms as in \cite{Yan2020_mmrapid}. Although many existing \ac{BT} algorithms rely on adaptive sensing codebooks based on prior estimates of channel parameters, e.g. hierarchical searches \cite{Noh2017_hierarchical}, we avoid these designs to reduce the beam management latency. Adaptive codebooks require subsequent sounding beams to be computed before the beamforming hardware can be updated, increasing the delay between measurements and the overall \ac{BA} overhead.

Let $\mathbf{w}_m\in \mathbb{C}^{N_r}$ be the $m$-th \ac{PN} sensing codeword used by the \ac{UE}. Assuming a 2-bit resolution of phase shifters in the \ac{UE}'s array, the $n$-th element of $\mathbf{w}_m$ is defined as $[\mathbf{w}_m]_n = e^{j\theta_n}$, where $\theta_n$ is the phase randomly selected from the set $\{0, \pi/2, \pi, 3\pi/2\}$. Thus, the \ac{PN} sensing codebook of $M$ codewords can be represented as the matrix $\mathbf{W}=[\mathbf{w}_1,...,\mathbf{w}_M] \in \mathbb{C}^{N_r \times M}$. Note that, for \ac{UE}s with phased arrays, the $M$ codewords in $\mathbf{W}$ are sequentially used for sensing. In this work, we assume that the entire sensing period is short and that the channel does not change during that time. Therefore, with any $\mathbf{w}_m$, the phase-less measurement $y_m^{(t)}$ at time step $t$ can be expressed as follows
\begin{align}
    y_m^{(t)} = \left|\mathbf{w}_m^H\mathbf{h}^{(t)} s_m^{(t)} + 
    \mathbf{w}_m^H\mathbf{n}_m^{(t)}\right|,
    \label{eq:pn_measurements}
\end{align}
where $s_m^{(t)}$ is the $m$-th transmitted symbol and $\mathbf{n}_m^{(t)}\sim\mathcal{CN}(0, \sigma^2\mathbf{I}_{N_r})$ is a complex vector of additive Gaussian noise. After vectorization, the $M$ measurements can be represented as $\mathbf{y}^{(t)}_{M}=[y_1^{(t)},...,y_M^{(t)}]^T$.

\subsection{Problem Statement}
Let $\mathbf{V} = [\mathbf{v}_1, ..., \mathbf{v}_K]\in \mathbb{C}^{N_r \times K}$ be a codebook of narrow pencil beams for data communication at the \ac{UE} side, where the $n$-th element of the $k$-th code is  $[\mathbf{v}_k]_n = \exp{(-j2\pi(n-1)\sin{\left(\pi (k-1)/K - \pi/2\right)}d/\lambda)}$. As is common in implementation, $K > N_r$ for high resolution \ac{AoA} estimates.
In each time step $t$, the goal of \ac{BT} is to keep track of the optimal beam $\mathbf{v}_{k^{(t)}}$ that results in the highest received signal power $z\left(k^{(t)}, t\right) = |\mathbf{v}_{k^{(t)}}^H \mathbf{h}^{(t)}s_m^{(t)}|^2$. However, in the presence of noise, only an estimate $\hat{k}^{(t)}$ of the optimal beam index $k^{(t)}$ is available, and it can be obtained through the exhaustive sweeping of all beams in $\mathbf{V}$. Mathematically, the estimate $\hat{k}^{(t)}$ is expressed as:
\begin{equation}
    \hat{k}^{(t)} = \argmax_k \left|\mathbf{v}_k^H\mathbf{h}^{(t)} s_k^{(t)} + 
    \mathbf{v}_k^H\mathbf{n}_k^{(t)}\right|.
\end{equation}
The ultimate goal of \ac{BP} based on phase-less measurements is to estimate $P$ beam indices $\hat{\mathbf{k}}^{(t)} = [\hat{k}^{(t)},\hat{k}^{(t+1)},...,\hat{k}^{(t+P-1)}]^T$ using a set of measurements $\mathcal{Y}^{(t-1)} = \{\mathbf{y}^{(t-1)} \in \mathbb{C}^{M_{I}}, \mathbf{y}^{(t-2)} \in  \mathbb{C}^{M_{L}},...,\mathbf{y}^{(t-T)}\in  \mathbb{C}^{M_{L}}\}$ made over the last $T$ time steps, where $t$ is the time step when the first prediction is required. Specifically, we aim to estimate $\widetilde{\mathbf{k}}^{(t)} = \argmax_{\mathbf{k}} \mathbb{P}(\mathbf{k} = \hat{\mathbf{k}}^{(t)}| \mathcal{Y}^{(t-1)})$. However, since the true probability is difficult to calculate, a \ac{BP} algorithm $p(\mathcal{Y}^{(t-1)})$ is proposed to approximate $\widetilde{\mathbf{k}}^{(t)}$ as follows:
\begin{equation}
    \widetilde{\mathbf{k}}^{(t)} = \argmax_{\mathbf{k}} \mathbb{P}(\mathbf{k} = \hat{\mathbf{k}}^{(t)}|\mathcal{Y}^{(t-1)}) \approx p(\mathcal{Y}^{(t-1)}).
    \label{eq:bt_approx}
\end{equation}

When designing $p(\mathcal{Y}^{(t-1)})$, accuracy, gain loss, and number of required measurements serve as performance metrics. Since \eqref{eq:bt_approx} represents a classification problem, the \textit{accuracy} of $p(\mathcal{Y}^{(t-1)})$ for $P$ test time steps is the fraction of time steps with the correct predicted beam direction: $acc(\mathbf{\hat{k}}^{(t)},  p(\mathcal{Y}^{(t-1)})) = \frac{1}{P}\sum_{j = 1}^P\mathbbm{1}[\hat{k}^{(j)} = p(\mathcal{Y}^{(t-1)})^{(j)}]$. Accuracy, however, does not capture the impact of incorrect prediction on data communications while tracking. In this work, the alignment quality is measured using \textit{gain loss} $g = z(k^{(t)}, t)/z(\hat{k}^{(t)}, t)$ or the difference in gain (in dB) between the optimal pencil beam and the pencil beam selected by the \ac{BT} and \ac{BP} algorithm, for a specified percentile of $P$ estimates. To evaluate the effective reduction in overhead, this work uses the \textit{average required number of measurements} over a tracking period to meet a maximum gain loss requirement. 

The goal of this work is to propose and demonstrate a machine learning based algorithm $p(\mathcal{Y}^{(t-1)})$ that uses phase-less measurements from a fixed set of \ac{PN} beams to reduce the required number of measurements for \ac{mmW} \ac{BT}. Two state-of-the-art phase-less \ac{BA} act as performance baselines, namely a traditional exhaustive search of all pencil beams and the \ac{MLP}-based mmRAPID algorithm\cite{Yan2020_mmrapid}.

\section{Algorithm Design}
\label{sec:algo_design}
While the baseline phase-less \ac{BA} algorithms focus solely on reductions in the number of compressive measurements required for a single time step, \ac{BT} algorithms can take advantage of the temporal relationship between \ac{UE} motion, \ac{AoA}s, and measurements. In this work, we propose that the relationship over time between pencil beam indices and measurements can be used to predict future beam angles without additional measurements. This section presents our \ac{ML} algorithm for \ac{BT} and \ac{BP} to predict current and future beam indices from current and past compressive measurements. 

\begin{figure}[t]
    \vspace{0mm}
    \centering
    \includegraphics[width = 1.0\linewidth,trim=15 15 15 15,clip]{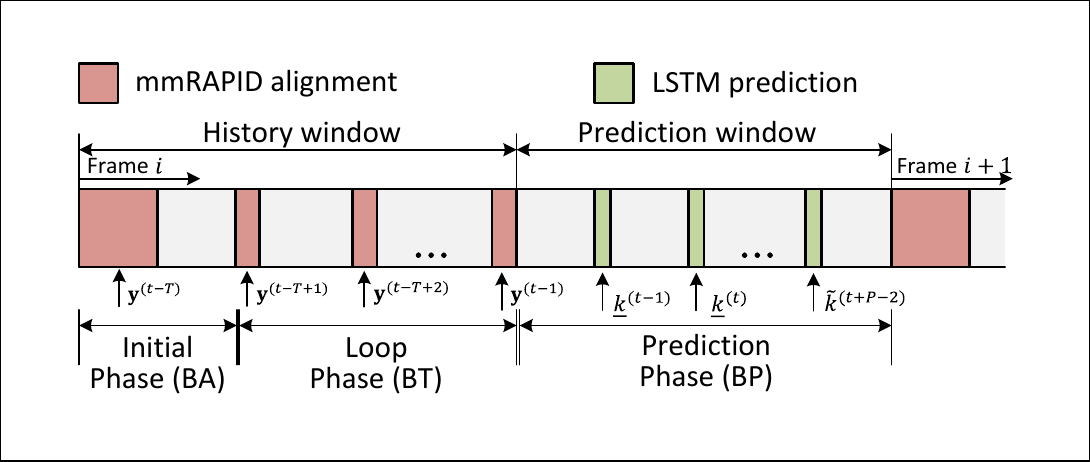}
    \vspace{-6mm}
    \caption{Proposed frame structure for \ac{BA}, \ac{BT}, and \ac{BP}}
    \label{fig:frame}
    \vspace{-6mm}
\end{figure}

Our proposed algorithm utilizes \ac{ML} instead of traditional compressive sensing or signal processing due to \ac{ML}'s ability to adapt to hardware impairments and potential for better beam index estimation. Poor synchronization and strong phase noise from \ac{mmW} oscillators can significantly affect measurement beams, leaving model-based techniques with poor estimation performance due to imperfect knowledge of the sensing dictionary. 
Also, \ac{MP} compressive sensing methods are known to be sub-optimal, with some \ac{ML} solutions providing better estimation than \ac{RSS}-\ac{MP} \cite{Yan2020_mmrapid}.

\subsection{Machine Learning Architecture}
\label{subsec:MLarch}
To naturally accommodate time series relationships, we selected an \ac{LSTM}-based \ac{Seq2Seq} algorithm for \ac{BP}. \ac{LSTM} networks are designed to handle feedback of learned features, incorporating the concept of memory and forgetting old information. Likewise, \ac{Seq2Seq} architectures are commonly used with \ac{LSTM} networks for machine translation in \ac{NLP} \cite{Sutskever2014_lstmSeq2Seq}. A \ac{Seq2Seq} algorithm for \ac{NLP} would use an \textit{encoder} to learn features sequentially from the words of an input language sentence and use a \textit{decoder} to predict the following translated word in the output language based on the previous output language word and the learned features from the encoder. 

The proposed algorithm borrows from these \ac{NLP} designs to predict future best pencil beam indices from the previous best beam index estimates and the previous phase-less measurements. This design repeats over non-overlapping frames of $T + P$ consecutive traditional \ac{BA} time steps, where $T$ is the number of steps in the \textit{history window} where measurements are taken and $P$ is the number of steps in the \textit{prediction window} where no measurements are taken. 
The frames can also be functionally described in three distinct periods conducting \ac{BA}, \ac{BT}, and \ac{BP} as described in Fig. \ref{fig:frame}. In the \textit{initial phase}, or the first time step, the \ac{UE} collects $M_I$ measurements for the first \ac{BA} using mmRAPID instance \#1. The \ac{UE} then collects $M_L$ measurements in the next $T-1$ time steps for the \textit{loop phase}, using mmRAPID instance \#2 for \ac{BA} as the buffer of measurements required to run the \ac{LSTM} encoder is filling. The beam index estimate from mmRAPID \#1 or \#2 at time step $t$ is denoted as $\underline{k}^{(t)}$. Intuitively, we restrict $M_I \geq M_L$, since the subsequent tracking and prediction likely relies most heavily on the initial beam estimate and more measurements generally leads to higher estimation accuracy. 
Note that the \ac{LSTM} encoder uses $M_L$ measurements at all time steps, including the first step where any additional $M_I - M_L$ measurements are only used for mmRAPID.
Finally, during the \textit{prediction phase} or the prediction window, no measurements are collected and the \ac{LSTM} is solely responsible for \ac{BP}. In total, this algorithm requires $M_I + M_L (T-1)$ total measurements to complete \ac{BA} and \ac{BT} for the frame, whereas a baseline \ac{BA} algorithm would require $M (T + P)$ measurements where $M$ is the required number of measurements for \ac{BA}.

\begin{figure*}[t]
    \vspace{0mm}
    \centering
    \includegraphics[width = 0.75\linewidth,trim=2 10 2 15,clip]{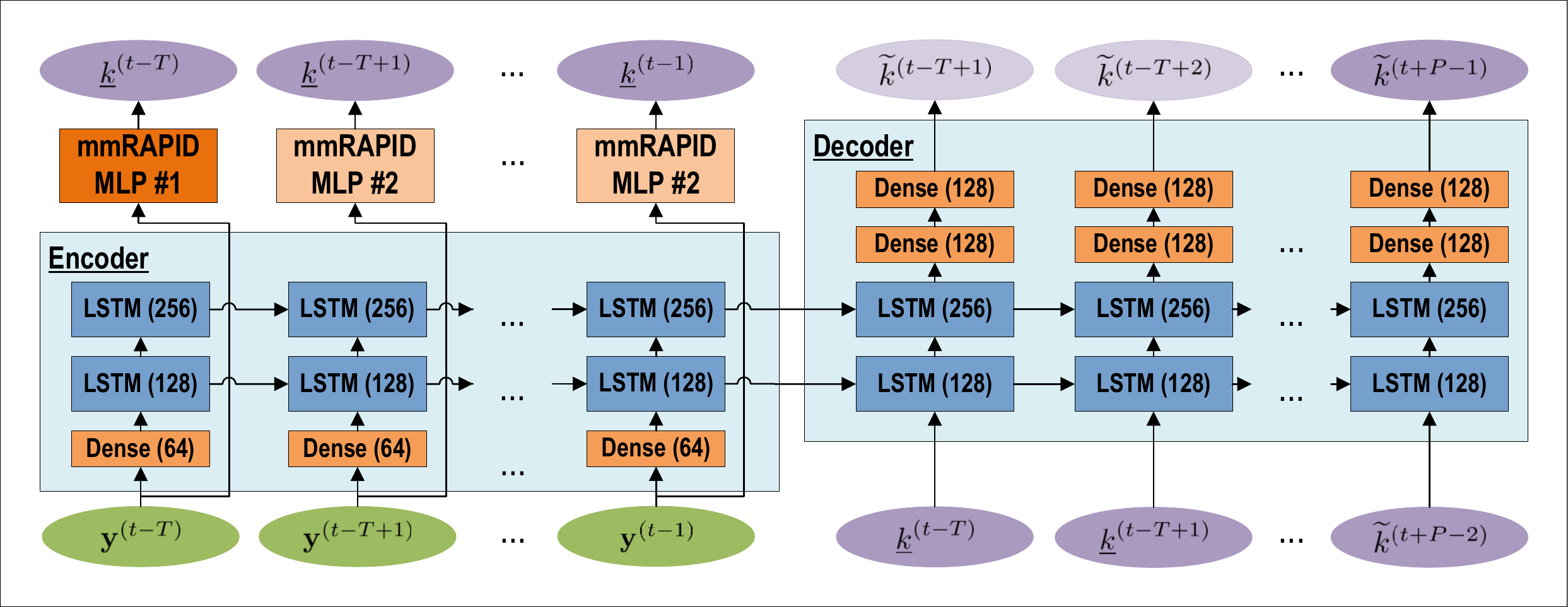}
    \vspace{-2mm}
    \caption{\ac{Seq2Seq} \ac{LSTM} machine learning architecture unrolled over time. Dropout and batch norm layers are not shown.}
    \label{fig:architecture}
    \vspace{-3mm}
\end{figure*}

Our \ac{Seq2Seq} architecture is shown in Fig. \ref{fig:architecture}, unrolled over time for clarity. Each block in the \ac{Seq2Seq} network is labeled with its hidden dimension in parentheses. To improve the training performance, each dense layer has a 10\% dropout layer and a batch norm layer at their output and the encoder input features are normalized to have a unit 2-norm. Additionally, the decoder beam index estimates, denoted as $\widetilde{k}^{(t)}$, are used as the predictions for $P$ time steps $t,...,t+P-1$ but are discarded for the $T-1$ time steps $t-T+1,...,t-1$. The decoder still provides these discarded estimates, since the decoder \ac{LSTM} may be able to learn from the sequence of prior beam  estimates to improve later predictions. During training time, both learning algorithms use the sparse categorical cross entropy loss function, with the RMSprop and Adam as optimizers for the mmRAPIDs and the LSTMs respectively. 
Both were developed in Tensorflow using the Keras API.

\subsection{Three-Stage Data-Driven Phase-less Beam Tracking and Prediction}
\label{subsec:ml3stage}

During the lifetime of a \ac{UE}, the proposed algorithm operates in three major stages for training and operation, based on the two-stage method in \cite{Yan2020_mmrapid}. During \textit{Stage 1}, the algorithm gathers training points for both mmRAPIDs and the \ac{LSTM} by conducting both an exhaustive search using the directional codebook $V$ and the compressive measurements with $M_I$ codewords from $W$. Once enough distinct \ac{AoA}s have been collected to train mmRAPID, the \ac{UE} starts \textit{Stage 2} and conducts the exhaustive search and $M_L$ compressive codewords, solely for training the \ac{LSTM}. mmRAPID and \ac{LSTM} training can take place locally on the \ac{UE} device or remotely in the cloud, with measurement data and labels uploaded for training on a server to save \ac{UE} power. Finally, during \textit{Stage 3}, the \ac{UE} can use the trained mmRAPID and LSTM to predict frames of angles, reducing the overall measurement requirements.

\begin{figure*}[t]
    \centering
    \begin{minipage}{0.41\textwidth}
    \centering  
    \includegraphics[width=\textwidth,trim=0 3 0 5,clip]{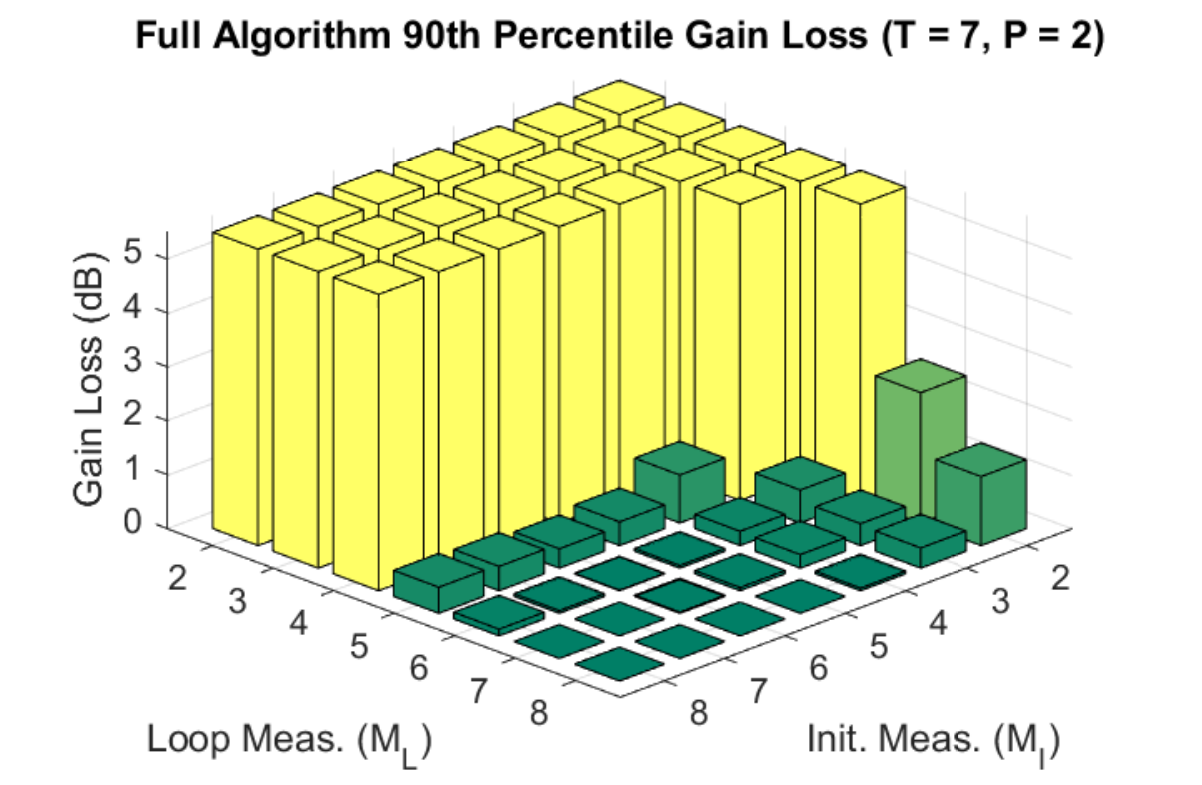}
    \small (a) Gain loss of the proposed algorithm
    \end{minipage}%
    \hspace{1cm}
    \begin{minipage}{0.41\textwidth}
    \centering  
    \includegraphics[width=\textwidth,trim=0 3 0 5,clip]{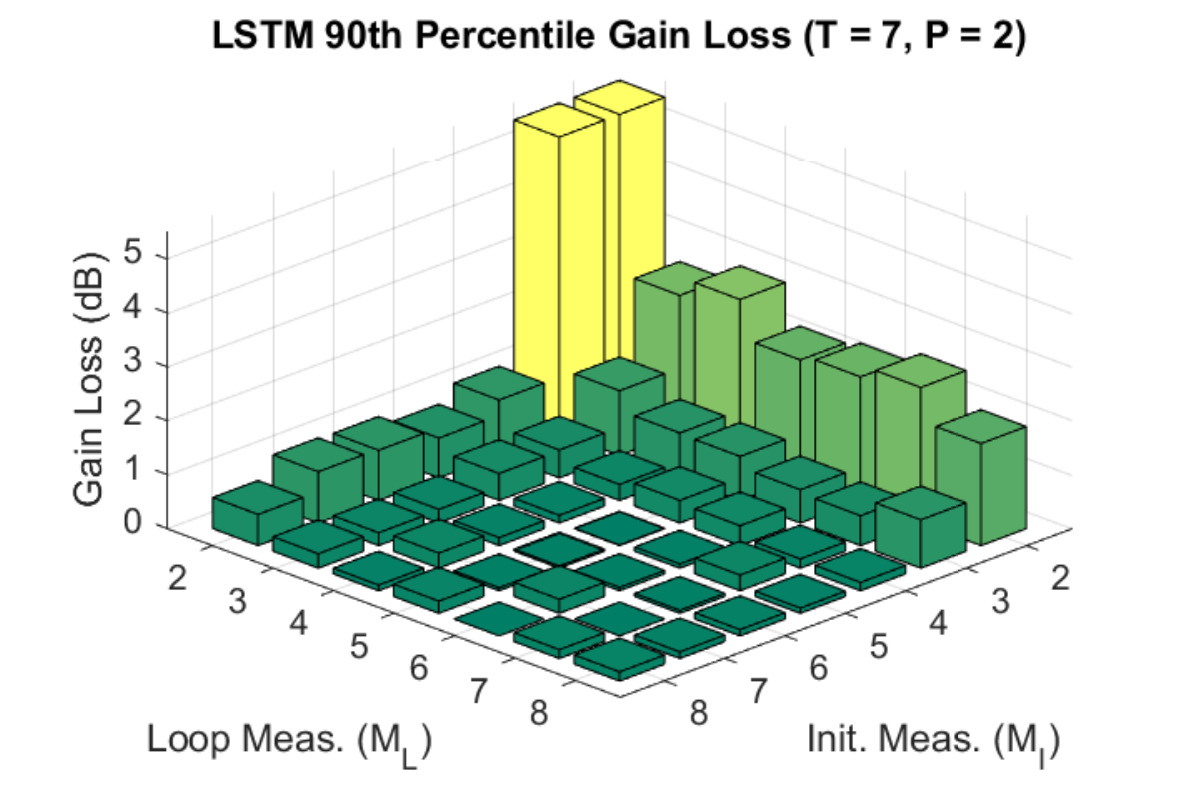}
    \small (b) Gain loss of the algorithm with LSTM predictions
    \end{minipage}%
    \caption{90th percentile gain loss over varied number of measurements and fixed window lengths $T = 7, P = 2$.}
    \label{fig:meas}
    \vspace{-6mm}
\end{figure*}

\section{Algorithm Evaluation}
\label{sec:eval_design}

\subsection{Simulation Design and Data Preparation}
\label{subsec:sim_design}
To test our proposed algorithm, we used 60 GHz channel data from the DeepMIMO dataset \cite{Alkhateeb2019_deepmimo}, based on ray-tracing simulations in Wireless Insite \cite{Remcom}. We selected the O1 scenario, an outdoor urban environment, for \ac{LOS} channels from \ac{UE}s with $N_r = 36$, $d=\frac{\lambda}{2}$, and 128 pencil beams in $V$ to a \ac{BS} at location 1. Since we assume the \ac{BS} completes \ac{BT} before the \ac{UE} with high \ac{AoA} resolution to maintain consistent beamforming gain, we emulate the \ac{BS} array with a single omnidirectional antenna. 
In practice, the proposed algorithm can reduce the overhead of a joint \ac{BS}-\ac{UE} exhaustive scan by repeatedly replacing the \ac{UE} exhaustive search with each \ac{BS} sector beam transmission.
Also, the \ac{UE} array's broadside is always perpendicular to the street and the motion of the \ac{UE}.

\ac{UE} trajectories were generated from the DeepMIMO dataset by emulating movement over the 0.2 m grid of available channel measurement locations. For this work, we selected \ac{UE} locations within the dataset's longest street (i.e. from UG 1). For each trajectory, our simulations selected a  random initial starting position near the right end of the street as well as a random constant velocity, both uniformly distributed over a 26 m by 36 m bounding box and integer velocities between 1 and 10 grid steps (0.2 m) per time step respectively. Simulations then selected positions straight down the street to the left, along the y dimension, on the \ac{UE} position grid. The measurement frames were generated by collecting all shifted windows within parts of trajectories within a "cell" area near the selected \ac{BS}, a rectangle 20 m across the street and 40 m along the street. The "cell" area was placed 15 m away from the \ac{BS} to limit the input angles to around $\pm 53^{\circ}$, a realistic restriction for antenna arrays. Our dataset's 30 dB maximum SNR for the beam measurements was constrained by the nearest \ac{UE} position to the \ac{BS}. With equivalent \ac{BS} transmit and noise power, all other \ac{UE} positions see lower \ac{SNR} with an average of 15 dB. 

In the final dataset, a total of 1000 trajectories were generated. This dataset was randomly split with 10\%, 75\%, and 15\% used for Stage 1 training, Stage 2 training, and testing sets respectively. 
Note that 10\% of the \ac{LSTM} training data, or 8.5\% of the total dataset, was used to validate the \ac{LSTM} architecture and hyperparameters.

\subsection{Simulation Results}
\label{subsec:results}

We first compare the 90th percentile gain loss of the proposed algorithm over varied number of initialization measurements $M_I$ and loop measurements $M_L$ with fixed window lengths in Fig. \ref{fig:meas}. Note that in each of these 3D bar plots, the yellow bars taller than 5 dB have been cutoff for readability and actually exhibit much higher gain loss. The "full algorithm" results in Fig. \ref{fig:meas}(a) show the combined performance of \ac{BA}, \ac{BT}, and \ac{BP} as described in Section \ref{sec:algo_design}, while the "LSTM estimates" in Fig. \ref{fig:meas}(b) show performance of initial \ac{BA} and the \ac{LSTM}'s predictions over the rest of the frame. With $T=7, P=2$, the results in Fig. \ref{fig:meas}(a) are dominated by mmRAPID \#2's estimates, demonstrating that the performance is driven by $M_L$. While larger $M_I$ does reduce gain loss after $M_L=5$, the required number of measurements for mmRAPID as found in separate simulations with the same dataset, the reduction is minimal and does not warrant $M_I > M_L$.
Fig. \ref{fig:meas}(b) supports the \ac{Seq2Seq} architecture's prediction capability, as the LSTM prediction gain loss is consistently low for nearly all $M_L$. In fact, the LSTM has lower gain loss at small $M_L$ than the full algorithm, demonstrating that the LSTM is taking advantage of the history window. 

\begin{figure*}[t]
    \vspace{0.032in}
    \centering
    \begin{minipage}{0.41\textwidth}
    \centering  
    \includegraphics[width=\textwidth,trim=0 3 0 5,clip]{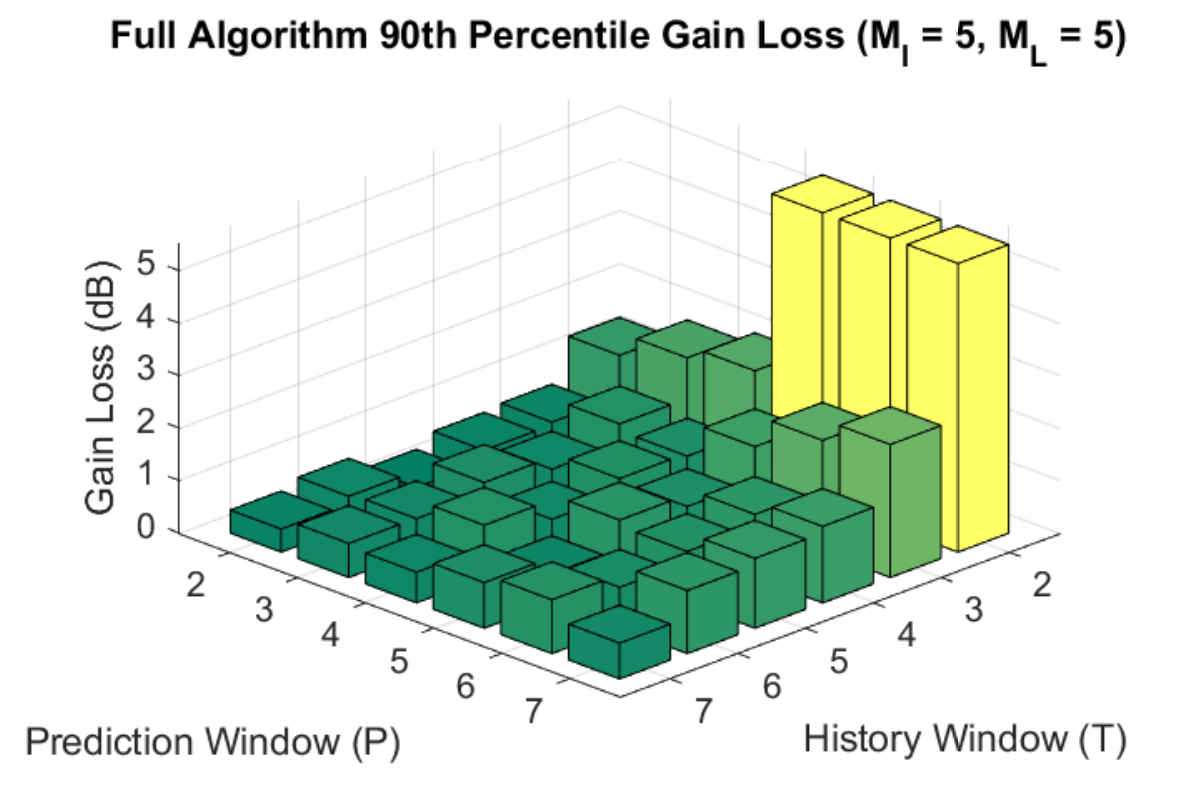}
    \small (a) Gain loss of the proposed algorithm
    \end{minipage}%
    \hspace{1cm}
    \begin{minipage}{0.41\textwidth}
    \centering  
    \includegraphics[width=\textwidth,trim=0 3 0 5,clip]{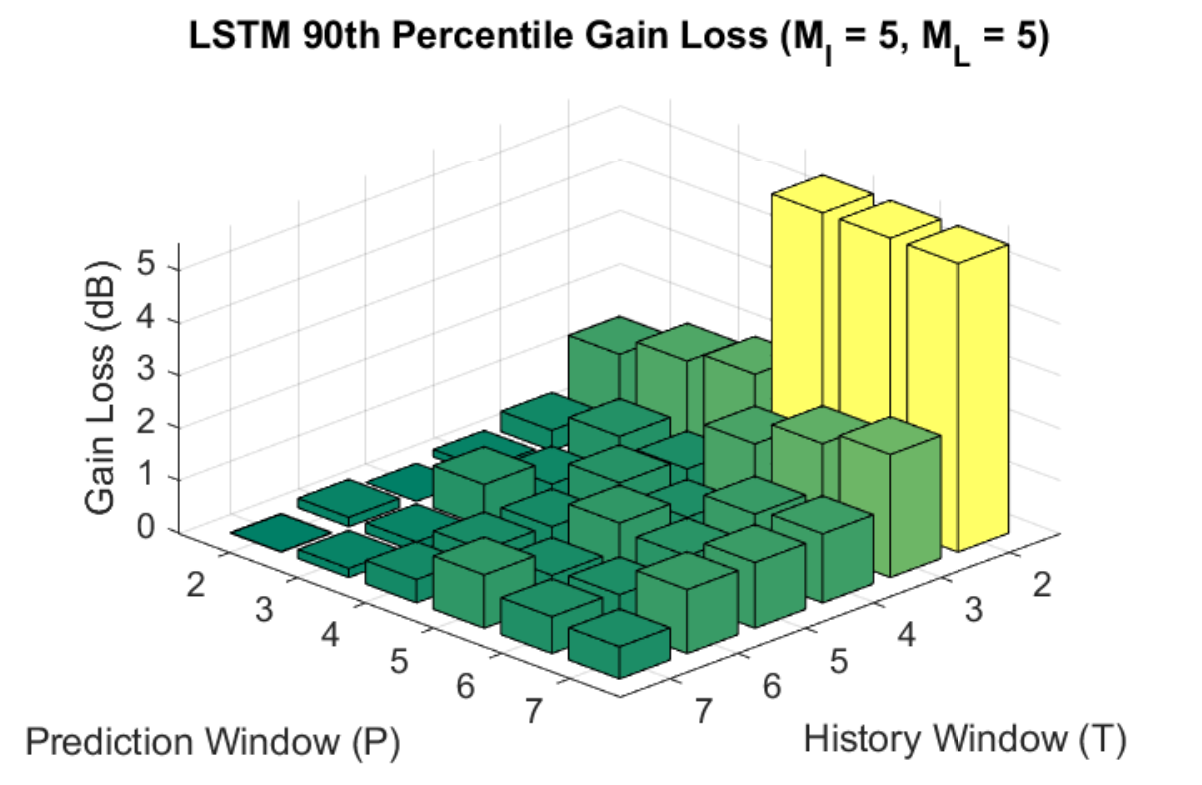}
    \small (b) Gain loss of the algorithm with LSTM predictions
    \end{minipage}%
    \caption{90th percentile gain loss over varied window sizes and fixed number of measurements $M_I = 7, M_L = 4$.}
    \label{fig:window}
    \vspace{-6mm}
\end{figure*}

Fig. \ref{fig:window} then presents the impact of the prediction and history window lengths on the gain loss. Both plots generally show increased gain loss for larger $P$ and smaller $T$, an intuitive result as the former requires the algorithm to extrapolate more predictions from the same data and the latter requires the algorithm to extrapolate the same predictions from less data. 
The gain loss varies between combinations of $P$ and $T$ due to the stochastic nature of the algorithm training. 
With large $T$ and small $P$, the \ac{LSTM} estimates show even lower gain loss than the full algorithm. Fig. \ref{fig:window}(b) demonstrates that, for a given number of measurements and enough historical data, the \ac{LSTM} can provide better estimates than mmRAPID.

When comparing the average required number of measurements over beam tracking frames, we find that the proposed algorithm significantly decreases the beam alignment overhead with 90th percentile gain loss of 3 dB or less. While the exhaustive search and mmRAPID require $M=128$ and $M=5$ measurements per time step respectively, the proposed algorithm requires only an average of 1.82 measurements per time step - a 98.6\% or 63.6\% reduction in overhead.
\footnote{Overhead reduction = $1 - (\left(M_I + M_L(T-1)\right)/\left(M(T+P)\right))$}
This was computed from the experiment with $M_I = 5, M_L = 5, T = 4, P = 7$, which used only 20 measurements for 11 time steps and observed only 1.5 dB 90th percentile gain loss.

\section{Conclusion and Future Work}
\label{sec:conclusion}
In this paper, we propose a novel phase-less, fixed-codebook \ac{UE} beam tracking and prediction algorithm. Through \ac{LOS} ray-tracing simulations, we demonstrate the \ac{LSTM} \ac{Seq2Seq} architecture is capable of predicting future beam directions and can save 98.6\% of the beam alignment overhead compared to an exhaustive search and 63.6\% compared to mmRAPID, a state-of-the-art \ac{ML} \ac{BA} algorithm. Several open questions remain with \ac{Seq2Seq} phase-less \ac{BT} and \ac{BP} algorithms. Multipath channels would likely impact prediction performance, as with mmRAPID \cite{Domae2021_globecom}, and would correspond to the algorithm's ability to generalize to new environments. \ac{BP} with more complex \ac{UE} trajectories, alternative sounding codebooks, and joint \ac{BS}-\ac{UE} \ac{BA} have also not yet been explored.

\bibliographystyle{IEEEtran}
\bibliography{IEEEabrv,references}

\end{document}